# A novel quasigroup substitution scheme for chaos based image encryption

Vinod Patidar[1,*], G. Purohit[1] and N. K. Pareek[2]

[1]*Department of Physics, School of Engineering, Sir Padampat SInghania University, Bhatewar, Udaipur – 313601, Rajasthan, INDI A*

[2]*University Computer Centre, Vigyan Bhawan, M.L.S. University, Udaipur -313001, Rajasthan, INDIA*

*vinod.patidar@spsu.ac.in, patidar.vinod@gmail.com

# A novel quasigroup substitution scheme for chaos based image encryption

During last two decades, there has been a prolific growth in the chaos based image encryption algorithms. Up to an extent these algorithms have been able to provide an alternative to exchange large media files (images and videos) over the networks in a secure way. However, there have been some issues with the implementation of chaos based image ciphers in practice. One of them is reduced/small key space due to the fact that chaotic behavior is only observed for certain range of system parameters/initial conditions of the chaotic system used in such algorithms. To overcome this difficulty, we propose a simple, efficient and robust image encryption algorithm based on combined applications of quasigroups and chaotic standard map. The proposed image cipher is based on the Shannon's popular substitution-diffusion architecture where a quasigroup of order 256 and chaotic standard map have been used for the substitution and permutation of image pixels respectively. Due to the introduction of quasigroup as part of the secret key along with the parameter and initial conditions of the chaotic standard map, the key space has been increased significantly. The proposed image cipher is very fast due to the fact that the substitution based on the quasigroup operations is very simple and can be executed easily through the lookup table operations on Latin squares (which are Cayley operation tables of quasigroups) and the permutation is performed row-by-row as well as column-by-column using the pseudo random number sequences generated through the chaotic standard map. The security and performance have been analyzed through the histograms, correlation coefficients, information entropy, key sensitivity analysis, differential analysis, key space analysis etc. and the results prove the efficiency and robustness of the proposed image cipher against the possible security threats.

## 1. Introduction

Chaos based image encryption has been an active research area over the last few years which has proven that a suitable choice of chaotic systems along with an intelligently designed combination of mixing transformations (substitution and diffusion transformations) comprising of complex involvement of the key leads to very fast and potentially reliable image encryption algorithm [Fridrich, 1998, Mao *et al.*, 2004, Chen *et al.*, 2004, Pareek *et al.*, 2006, Patidar *et al.*, 2009]. Since the dynamics of the chaotic system, used in most of the chaos based cryptosystems, is control/system parameter dependent, they exhibit desired chaotic behavior only for certain ranges/values of control/system parameters. Hence the cryptosystems based on chaotic systems mainly suffer from key size requirements [Alvarez & Li, 2006, Kelber & Schwarz, 2007, Rhouma *et al.*, 2010, Solak *et al.*, 2010, Ozkaynak and Ozer, 2016]. Despite above mentioned drawback of having a reduced key space, the chaotic systems have been a preferred choice [Patidar *et al.*, 2011, Wang and Xu, 2015, Wu *et al.*, 2016, Abd-El-Hafiz *et al.*, 2016, Farajallah *et al.*, 2016] for the image encryption systems due to their features: ergodicity, mixing property, sensitivity to initial conditions/system parameters, etc. having close resemblance with the ideal cryptographic properties: confusion, diffusion, balance, avalanche properties, etc. [Alvarez & Li, 2006].

The algebraic structures known as Quasigroups [Pflugfelder, 1991] also provide a powerful method for generating a larger set of permutation/substitution transformations by rearranging not only the data values, but also changing the data values across their range. The Cayley table of a finite quasi group of order n is a Latin square [Dénes & Keedwell, 1991] and the number of Latin squares of order n increases very quickly with n and indeed very large for rather small n [Lint & Wilson, 1992, Jacobson & Matthews, 1996]. Hence, such transformations can provide a potent combination of confusion and diffusion in the encryption of digital images and also the Quasigroup based cryptosystem can generate more number of keys even when we consider a small order quasigroup.

In most of the conventional cryptographic algorithms [Schneier, 1996] the associative algebraic structures such as groups, rings and fields are used. Dénes and Keedwell [Dénes & Keedwell, 1992, Keedwell 1999] proclaimed the start of new era in cryptography by using the non-associative algebraic structures such as quasigroups and neo-fields. Since then there have been some attempts of designing quasigroup based pseudorandom number generators/cryptographic schemes [Kościelny, 1996, Kościelny & Mullen, 1999, Golomb *et al.* 2001, Gligoroski, 2004, Gligoroski, 2005, Markovski *et al.*, 2005, Satti and Kak, 2009, Battey and Parakh, 2013] but weaknesses have been found in them [Dichtl & Böffgen, 2012] and have seen relatively less success than chaos based systems. Some of the comprehensive and in-depth overviews of the recent developments and current state of the art in the field of applications of quasigrous in cryptology can be seen in [Shcherbacov, 2012, Mileva, 2014].

As mentioned earlier that the chaos based cryptosystems mainly suffer from key size requirements and the quasigroups can provide a larger set of permutation/substitution transformations i.e., an immensely large number of keys. Keeping these facts in mind we propose in this paper a simple, efficient and robust image encryption algorithm based on combined applications of quasigroups and chaotic dynamical systems. The proposed image cipher is a block cipher and based on the Shannon's substitution-diffusion architecture [Shannon, 1949] which has been the main core structure of most of the chaos based image encryption algorithms developed during last one and half decade. In the proposed image cipher we use a quasigroup of order 256 and chaotic standard map for the substitution and permutation of image pixels respectively. The secret key of the image cipher comprises of a quasigroup of order 256, initial conditions & parameter of chaotic standard map. The encryption in the proposed image cipher has minimum (maximum) two (sixteen) rounds and each round has a substitution and a permutation module. The substitution is done column-wise and row-wise starting from the first and last pixel respectively through the left and right quasigroup binary operations on a 256 order quasigroup (which is part of the secret key). However the permutation is performed row-by-row as well as column-by-column using the pseudo random number sequences generated through the chaotic standard map. Due to introduction of quasigroup as the part of secret key, the proposed image cipher does not suffer with the key size requirement (as opposed to the most of the existing chaos based image ciphers) since the no. of possible quasigroups of order 256 lies somewhere in the range $0.753 \times 10^{102805} \geq L(256) \geq 0.304 \times 10^{101724}$ [Lint & Wilson, 1992, Jacobson & Matthews, 1996]. The proposed image cipher is very fast due to the fact that the substitution based on the quasigroup operations are simple and can be executed easily through the lookup table operations on Latin squares (which are Cayley operation tables of quasigroups) without any explicit binary quasigroup operation calculation. We also present the results of our security and performance analysis of the proposed image encryption done through the histograms, correlation coefficients, information entropy, key sensitivity analysis, differential analysis, key space analysis etc. Results are encouraging and illustrate the efficiency and robustness of the proposed image cipher against the possible security threats. The rest of the paper is organized as follows: In Section 2, we briefly introduce the notion of quasigroup and associated operations which have been used in the proposed algorithm. In Section 3 we describe in detail the complete process of encryption and decryption used in the proposed image cipher. In Sections 4 to 7, we summarize the results of our statistical, key-sensitivity, differential and key space analyses to assess the performance and robustness of the proposed image cipher. Finally Section 8 concludes the paper.

**2. The quasigroup**

The structure $(Q, \circledast)$, $Q = \{q_1, q_2, q_3, q_4, \ldots \ldots \ldots, q_n\}$, $\|Q\| = n$ is called a finite quasigroup of order n [Pflugfelder, 1991] if,

$\forall\ a, b\ \in Q, \exists\ x, y \in Q$ such that $a \circledast x = b$ and $y \circledast a = b$

$\forall\ x,\ y,\ z \in Q,\ x \circledast y = x \circledast z \Rightarrow y = z$ and $y \circledast x = z \circledast x \Rightarrow y = z$

Thus the Cayley table of a finite quasigroup of order n is a Latin square of order n and vice-versa. A Latin square is an $n \times n$ array filled with n different symbols in such a way that each of the symbol occurring exactly once in each row and exactly once in each column [Dénes & Keedwell, 1991]. With every quasigroup $(Q, \circledast)$ there exist two more operations $\oslash$ and $\obslash$ satisfying the following

$\forall\ x, y, z \in Q$ such that $x \circledast y = z \Leftrightarrow y = x \oslash z \Leftrightarrow x = z \obslash y$.

These operations $\oslash$ and $\obslash$ can be referred as left and right inverse quasigroup operations respectively and the structures $(Q, \oslash)$ and $(Q, \obslash)$ also form quasigroups of the same order. In Figure 1, we have given example of an order 5 quasigroup along with its left and right inverse.

The number of Latin squares of order $n$ increases very quickly with $n$ and indeed very large for rather small $n$. The total number of all possible Latin squares $L(n)$ of order $n \times n$ is only known up to $n = 11$, which is L(11)= 776966836171770144107444346734230682311065600000. The most accurate upper and lower bound for the number of Latin squares of order $n$ is [Lint & Wilson, 1992, Jacobson & Matthews, 1996]

$$\prod_{k=1}^{n}(k!)^{n/k} \geq L(n) \geq \frac{(n!)^{2n}}{n^{n^2}}$$

According to this, the upper and lower bounds for the $L(256)$ are: $0.753 \times 10^{102805} \geq L(256) \geq 0.304 \times 10^{101724}$. Therefore we may safely conclude that there exist atleast $n!\,(n-1)!\,....2!\,1!$ Latin squares of order $n$. Hence such Latin squares can be potentially used as secret keys in symmetric cryptography.

## 3. The proposed image cipher

### 3.1 The original image to be encrypted:

The original colour image of height $H$ and width $W$ is the image to be encrypted. The image is read as a 3D matrix $PI(i,j,k)$ of integers between 0 and 255, where $1 \le i \le H$, $1 \le j \le W$ and $1 \le k \le 3$. Thus, we have total $H \times W \times 3$ integer elements in the 3D input image matrix.

### 3.2 The secret key:

The secret key in the proposed image encryption technique is divided into two parts. First part consists of a set of three floating point numbers and four integers $(x_0, y_0, K, NS, NR, seed1, seed2)$, where $x_0, y_0 \in (0, 2\pi)$, $K$ can have any real value greater than 18.0, $NS$ is an integer value preferably greater than 100, $NR$ is an integer value between 1 and 16, which is treated as number of encryption rounds, $seed1$ and $seed2$ are integers between 0 to 255 and used as the seed values for the initiation of the substitution. The second part is a $256 \times 256$ gray image whose gray pixel values (+1) form a $256 \times 256$ Latin square which is the Cayley operation table of a quasigroup. This table is basically used as the lookup table for the quasigroup substitution. The generation of quasigroups of desired order through software or hardware is very easy [Kościelny, 2002, Mihajloska et al., 2013] hence the secret key of the proposed image encryption scheme can be generated very easily.

### 3.3 Calculation of the dimension of new 2D matrix:

Here we calculate the dimension of 2D matrix ($NH$ and $NW$) using the values of $W$ and $H$. It is calculated in such a way that $H \times W \times 3 = NH \times NW$ and $(NW - NH)$ is minimum, hence $NW \geq NH$. With the above mentioned conditions, we try to form a square 2D matrix as far as possible from all the elements of 3D image matrix, if it is not possible then we form a rectangular matrix with the lowest possible difference in the number of rows and columns. The $H \times W \times 3$ integer elements are then arranged into a 2D matrix P(i,j) where $1 \leq i \leq NH$ and $1 \leq j \leq NW$.

### 3.4 Substitution using the quasi group binary operations:

In the proposed image cipher, we use quasigroup operation table to introduce the confusion/substitution. As explained above in the secret key section that we use a quasigroup of order 256 i.e. its Cayley table is a 256X256 matrix consisting of integers ranging from 1 to 256 in such a way that no integer occurs twice in a particular row or column. However, the pixel intensity level in RGB images vary from 0 to 255 hence before starting the actual substitution/confusion, we add 1 to all the integers of the 2D matrix P(i, j).

*3.4.1 Column wise substitution starting from the first pixel (left quasigroup binary operation)*

```
for j=1 to NW
    for i=1 to NH
        if (i=1 & j=1)
            P(i,j)=seed1⊛P(i, j)
        else
            if (i=1 & j≠1)
                P(i, j)=P(NH, j-1)⊛P(i, j)
            else
                P(i,j)=P(i-1, j)⊛P(i, j)
            end
        end
    end
end
```

*3.4.2 Row wise substitution starting from the last pixel (right quasigroup binary operation)*

```
        for i = NH to 1
           for j=NW to 1
              if (j=NW &  i=NH)
                 P(i,j)=P(i, j)⊛seed2
              else
                 if (j=NW & i≠NH)
                    P(i, j)=P(i, j)⊛P(i+1, 1)
                 else
                    P(i,j)=P(i, j)⊛P(i, j+1)
                 end
              end
           end
        end
```

After completion of the quasigroup substitution, subtract 1 from all the integers of the 2D matrix.

### 3.5 Permutation of rows and columns using chaotic standard map:

The permutation of the pixels in the proposed image cipher is a simplified version of the chaotic standard map based permutation used in the pseudorandom-substitution scheme [Patidar *et al.*, 2011]. The execution/algorithmic details of the permutation used in the proposed image cipher are as follows:

Iterate the standard map *NS* times, starting with the initial conditions $(x_0, y_0)$ and using the parameter K specified in the secret key and the new values are stored as $(x, y)$.

Now calculate the number of iteration to skip from the 2D matrix obtained after the substitution in the following way.

$NSK = \{P(1,1) + P(1,2) + ... + P(1, NW) + P(2,1) + ... + P(NH, NW)\} \mod 256$

Then we generate the permutation boxes and then execute the permutation of rows and columns with the help of chaotic standard map in the following way:

```
        for k =1 to NSK
                x = (x + K sin y) mod 2π
                y = (y + x) mod 2π
        end
        for j =1 to NW step 1
                x = (x + K sin y) mod 2π
                y = (y + x) mod 2π
```
$$PR1(j) = 1 + \left\lfloor \frac{x}{2\pi} \cdot NH \right\rfloor$$

$$PC1(j) = 1 + \left\lfloor \frac{y}{2\pi} \cdot NW \right\rfloor$$

```
                x = (x + K sin y) mod 2π
                y = (y + x) mod 2π
```
$$PR2(j) = 1 + \left\lfloor \frac{x}{2\pi} \cdot NH \right\rfloor$$

$$PC2(j) = 1 + \left\lfloor \frac{y}{2\pi} \cdot NW \right\rfloor$$

```
        end
        for j =1 to NH step 1
                interchange P(PR1(j),:) and P(PR2(j),:)
        end
        for j =1 to NW step 1
                interchange P(:,PC1(j)) and P(:,PC2(j))
        end
        C(:, :)=P(:, :)
```

In the above algorithm $P(i,:)$, $P(:,j)$ and $P(:,:)$ respectively, represent all the elements of $i^{th}$ row, all the elements of $j^{th}$ column and all the elements of matrix P.

These modules of substitution and permutation are repeated sequentially $NR$ number of times (as specified in the secret key). We recommend at least two rounds of encryption to ensure the foolproof security. Finally the elements of 2D matrix $C(i, j)$ are arranged in a 3D matrix $CI(i, j, k)$ of dimension $H \times W \times 3$ followed by a conversion to a colour image. It is the final encrypted image. The whole process of encryption is depicted in a simple block diagram in Figure 2.

The process of decryption is completely reverse of the encryption process. For the proper decryption the same secret key is to be supplied and the encrypted image is processed in the same way as in the encryption. The only difference would be that the substitution and permutation operations will be executed in the reverse order. Below we describe the algorithms used for the recovery of permutation and substitution during the decryption process.

*3.6 Recover the permutation of rows and columns using chaotic standard map:*

Iterate the standard map $NS$ times, starting with the initial conditions $(x_0, y_0)$ and using the parameter $K$ specified in the secret key and the new values are stored as $(x, y)$.

Now calculate the number of iteration to skip from the 2D matrix $C(i, j)$ of the encrypted image (obtained in the same way as in encryption)

$$NSK = \{C(1,1) + C(1,2) + ... + C(1, NW) + C(2,1) + ... + C(NH, NW)\} \bmod 256$$

Then we generate the permutation boxes and then execute the permutation of rows and columns with the help of chaotic standard map in the following way:

```
for k =1 to NSK
        x = (x + K sin y) mod 2π
        y = (y + x) mod 2π
end
for j =1 to NW step 1
        x = (x + K sin y) mod 2π
        y = (y + x) mod 2π
        PR1(j) = 1 + ⌊ x/(2π) · NH ⌋
        PC1(j) = 1 + ⌊ y/(2π) · NW ⌋
        x = (x + K sin y) mod 2π
        y = (y + x) mod 2π
        PR2(j) = 1 + ⌊ x/(2π) · NH ⌋
        PC2(j) = 1 + ⌊ y/(2π) · NW ⌋
end
for j =NW to 1 step -1
        interchange C(:,PC2(j)) and C(:,PC1(j))
end
for j =NH to 1 step -1
        interchange C(PR2(j),:) and C(PR1(j),:)
end
```

### 3.7 Recover the substitution using inverse quasi group operations:

Before starting the recovery of substitution add 1 to all the elements of 2D matrix $C(i,j)$.

### 3.7.1 Recover the row wise substitution (right inverse quasigroup operation)

```
CRT(:, :)=C(:, :)
for i = NH to 1
   for j=NW to 1
      if (j=NW &  i=NH)
         C(i,j)=CRT(i, j)⊘seed2
      else
         if (j=NW & i≠NH)
            C(i, j)=CRT(i, j)⊘CRT(i+1, 1)
         else
            C(i,j)=CRT(i, j)⊘CRT(i, j+1)
         end
      end
   end
end
```

### 3.7.2 Recover the column wise substitution (left inverse quasigroup operation)

```
CCT(:, :) = C(:, :)
for j=1 to NW
   for i=1 to NH
      if (i=1 & j=1)
         C(i, j)=seed1⊘CCT(i, j)
      else
         if (i=1 & j≠1)
            C(i, j)=CCT(NH, j-1)⊘CCT(i, j)
         else
            C(i, j)=CCT(i-1, j)⊘CCT(i, j)
         end
      end
   end
end
P(:, :)=C(:, :)
```

After completion of the recovery of the substitution subtract 1 from all the integers of the output 2D matrix. In the above recovery algorithm ⊘ and ⊘ respectively, represent the right inverse and left inverse quasigroup operations. These recovery modules of substitution and permutation are repeated sequentially NR number of times (as specified in the secret key). Finally the elements of 2D matrix $P(i, j)$ are arranged in a 3D matrix $PI(i, j, k)$ of dimension $H \, X \, W \, X \, 3$ followed by a conversion to a colour image. It is the decrypted image.

## 4. The statistical analysis:

To check the statistical relationship between the plaintext and ciphertext produced using the proposed image cipher, we have done extensive statistical analysis by computing the image histograms, information entropy, correlation between the pair of plain and cipher images, and correlation between the adjacent pixels for a large number of images having widely different contents. In Table 1, we have shown the encryption of the two sample images i.e. image 'Lena' and an 'all-zero' image of size 200 X 200. Here the resultant images are shown after substitution using the quasigroup and permutation using the chaotic map for the two rounds. The secret key used for the encryption has also been mentioned in the first row of the table. In Figures 3 and 4 respectively, we have depicted the histograms of the plain and cipher images for the image 'Lena' and an 'all-zero' image for the first two rounds of encryption. Particularly in the first column-the histograms for the red, green and blue layers of the plain image have been shown however in the second and third columns the corresponding histograms of the cipher images after first and second rounds respectively have been depicted. It is very clear from the visual inspection of the histograms that the distribution of the pixel values in the cipher images is uniform/flat, random and independent of the original images.

To quantify the uniformity of the pixel distribution in the cipher images, we have also computed the information entropy / Shannon entropy [Shannon, 1949]. The information entropy quantifies the amount of information contained in data, usually in bits or bits/symbol. It is also the minimum message symbol length necessary to communicate the information. In other words, information entropy is a measure of disorder. For example, a long sequence of repeating characters has entropy 0, since every character is predictable and a truly random sequence has maximum entropy, since there is no way to predict the next character in the sequence.

The information entropy $H(m)$ of a message $m$ is calculated by using $H(m) = \sum_{i=0}^{2^N-1} P(m_i) \log_2 \frac{1}{P(m_i)}$ (bits), where $P(m_i)$ is the probability of occurrence of symbol $m_i$ in the message $m$ and $N$ is the number of bits required to represent a symbol in the message $m$. For a 24-bit colour image (I), the information entropy for each colour layer (Red, Green and Blue) is calculated using $H^{R/G/B}(I) = \sum_{i=0}^{2^8-1} P^{R/G/B}(I_i) \log_2 \frac{1}{P^{R/G/B}(I_i)}$ (bits).

A truly random image (RI) has uniform distribution of pixel intensities in the interval [0, 255] i.e. $P^{R/G/B}(RI_i) = 1/256$ for all $i \in [0, 255]$, hence $H^{R/G/B}(RI) = 8$ bits.

In Table 2, we have given our computed results of information entropy for the image 'Lena' and an 'all-zero' plain images and their corresponding cipher images produced using the secret keys given in Table 1. It is clear that the information entropy is larger than 7.99 for all the cipher images produced through the proposed image cipher.

We have also calculated the correlation between various pairs of plain and cipher images by computing the 2-dimensional correlation coefficients between various colour channels of the plain and cipher images. The 2D-correlation coefficients have been calculated using the following relation:

$$C_{AB} = \frac{\frac{1}{H \times W} \sum_{i=1}^{H} \sum_{j=1}^{W} (A_{i,j} - \overline{A})(B_{i,j} - \overline{B})}{\sqrt{\left(\frac{1}{H \times W} \sum_{i=1}^{H} \sum_{j=1}^{W} (A_{i,j} - \overline{A})^2\right)\left(\frac{1}{H \times W} \sum_{i=1}^{H} \sum_{j=1}^{W} (B_{i,j} - \overline{B})^2\right)}},$$

with $\overline{A} = \frac{1}{H \times W} \sum_{i=1}^{H} \sum_{j=1}^{W} A_{i,j}$ and $\overline{B} = \frac{1}{H \times W} \sum_{i=1}^{H} \sum_{j=1}^{W} B_{i,j}$.

Here $A$ represents one of the red ($R$), green ($G$) & blue ($B$) channels of the plain image, $B$ represents one of the red ($R$), green ($G$) & blue ($B$) channels of the cipher image, $\bar{A}$ & $\bar{B}$ respectively, are the mean values of the elements of 2D matrices $A$ & $B$ and $H$ & $W$ respectively, are the height & width of the plain/cipher image. The results of our calculation of correlation coefficients for the image 'Lena' and an 'all-zero' image and their corresponding cipher images (shown in Table 1) have been given in Table 3. It is clear that the correlation coefficients between various channels of the plain image and cipher image are very small (or practically zero), hence the cipher images bear no clue about their corresponding plain images.

To analyze the correlations of adjacent pixels (horizontally as well as vertically) in the plain and cipher images, we have also computed the correlation coefficients for all the pairs of horizontally and vertically adjacent pixels in all plain and cipher images. This computation has been done by using the following formula:

$$C = \frac{\frac{1}{N}\sum_{i=1}^{N}(x_i - \bar{x})(y_i - \bar{y})}{\sqrt{\left(\frac{1}{N}\sum_{i=1}^{N}(x_i - \bar{x})^2\right)\left(\frac{1}{N}\sum_{i=1}^{N}(y_i - \bar{y})^2\right)}},$$

with $\bar{x} = \frac{1}{N}\sum_{i=1}^{N}x_i$ and $\bar{y} = \frac{1}{N}\sum_{i=1}^{N}y_i$.

Here $x_i$ and $y_i$ form $i^{th}$ pair of horizontally/vertically adjacent pixels and $N$ is the total number of pairs of horizontally/vertically adjacent pixels. For an image of size $W \times H$ pixels $N = (W-1)H$ (for horizontally adjacent pixels) and $N = (H-1)W$ (for vertically adjacent pixels). The results of our calculation for the correlation coefficient between the horizontally and vertically adjacent pixels for the images 'Lena' and 'All-zero' and their corresponding cipher images are given in Table 4. We observe that adjacent pixels in plain images are highly correlated however for the pixels in cipher images the correlation is almost zero hence the proposed image encryption technique removes the correlation between the adjacent pixels.

## 5. The key sensitivity analysis:

Extreme key sensitivity guarantees the security of a cryptosystem against the brute force attacks. The key sensitivity means the cipher image produced by the cryptosystem should be very sensitive to the secret key i.e., if we use two slightly different keys to encrypt the same image then two cipher images produced should be completely independent of each other. To test the key sensitivity of the proposed image cipher, we have first encrypted the plain image 'Lena' with a specific choice of the secret key and store the encrypted image as 'Encrypted Image'. Now we make a small change in the one of the several parts of the secret key and produce the cipher image 'Encrypted Image X' and then we compare both the cipher images. The Table 5 gives the details of the various possibilities of minimal changes in the secret key and corresponding cipher images.

To quantifying mutual independence of these encrypted images we have calculated the 2D-correlation and mutual information for the pair of encrypted images. The 2D correlation has been calculated in the same manner as done in the statistical analysis section above. However the mutual information between the two encrypted images $I_1$ and $I_2$ is calculated using $MI(I_1; I_2) = H(I_1) + H(I_2) - H(I_1, I_2)$, here $H(I_1)$, $H(I_2)$ are the information entropies calculated for the image $I_1$ and $I_2$ using the probability distributions $P(I_1)$, $P(I_2)$ respectively. However $H(I_1, I_2)$ is the joint information entropy of $I_1$ and $I_2$ calculated using the joint probability distribution $P(I_1, I_2)$. In Table 6, we have given the computed results of the 2D correlation coefficients and mutual information for the pair of encrypted images produced through very similar keys (as described in Table 5). It is very clear that the correlation between the cipher images produced using the very similar keys is negligible and also the mutual information between them is very low indicating that knowing the content of one cipher image $I_1$ does not reveal any information about the other cipher image $I_2$ and vice versa. Hence the proposed image cipher exhibits very high key sensitivity.

## 6. The differential analysis

The differential analysis is the study of how small changes in the plaintext affect the resultant ciphertext with the application of the same secret key. It is generally done by implementing the chosen plaintext attack but now there are extensions which use known plaintext as well as ciphertext attacks also. For image cryptosystems a very common strategy to implement the differential analysis, is to make a slight change, usually one pixel, in the plain image and compare the two cipher images (obtained by applying the same key on two plain images having one pixel difference only). If a meaningful relationship can be between established the plain image and cipher image then it may further facilitates us in determining the secret key. However if the image cryptosystem produces significant, random and unpredictable changes in the ciphertext under one pixel change in the plain image then such differential analysis become inefficient. The two most common measures: NPCR (net pixel change rate) and UACI (unified average changing intensity) are used to test the vulnerability of the image cryptosystems against the differential analysis.

The NPCR is used to measure the percentage number of pixels in difference of a particular colour channel in two cipher images obtained by applying the same secret key on two plain images having one pixel difference only. If $C_{i,j}^{R/G/B}$ and $\overline{C}_{i,j}^{R/G/B}$ (where $1 \leq i \leq H$ and $1 \leq j \leq W$, $H$ is height and $W$ is width and R, G and B represent red, green and blue channels) represent two cipher images then NPCR for each colour channel is defined as:

$$NPCR^{R/G/B} = \frac{\sum_{i=1}^{H}\sum_{j=1}^{W} D_{i,j}^{R/G/B}}{W \times H} \times 100\%$$

where

$$D_{i,j}^{R/G/B} = \begin{cases} 0 & \text{if } C_{i,j}^{R/G/B} = \overline{C}_{i,j}^{R/G/B} \\ 1 & \text{if } C_{i,j}^{R/G/B} \neq \overline{C}_{i,j}^{R/G/B} \end{cases}.$$

The NPCR value for two random images, which is an expected estimate for an ideal image cryptosystem, is given by

$$NPCR_{Expected}^{R/G/B} = \left(1 - 2^{-L^{R/G/B}}\right) \times 100\%,$$

here $L^{R/G/B}$ is the number of bits used to represent the red, green or blue channels of the image. For a 24-bit true colour image (8 bit for each colour channel) $L^{R/G/B} = 8$ hence $NPCR_{Expected}^{R/G/B} = 99.6094\%$.

The UACI, the average intensity difference of a particular channel between two cipher images $C_{i,j}^{R/G/B}$ and $\overline{C}_{i,j}^{R/G/B}$, is defined as:

$$UACI^{R/G/B} = \frac{1}{W \times H} \sum_{i=1}^{H}\sum_{j=1}^{W} \frac{C_{i,j}^{R/G/B} - \overline{C}_{i,j}^{R/G/B}}{2^{L^{R/G/B}} - 1} \times 100\%.$$

The UACI value for two random images, which is an expected estimate for an ideal image cryptosystem is given by

$$UACI_{Expected}^{R/G/B} = \frac{1}{2^{2L^{R/G/B}}} \cdot \frac{\sum_{i=1}^{2^{L^{R/G/B}}-1} i(i+1)}{2^{L^{R/G/B}} - 1} \times 100\%.$$

For a 24-bit true colour image $UACI_{Expected}^{R/G/B} = 33.4635\%$.

We have done an extensive analysis to compute the NPCR and UACI for the proposed image cipher using two plain images 'Lena' and 'All-zero" and the secret key given in the first row of the Table 1. Particularly we have randomly chosen 200 different pixels (one at a time, including the very first and very last pixels of the image) in each plain image and changed one of the R, G, B intensity values by one unit only and computed the NPCR and UACI as explained above for all 200 cases each for both the plain images. The computed results of NPCR and UACI have been depicted in Figures 5 and 6 respectively for the 'Lena' and 'all-zero' images. It is clear that the NPCR and UACI values are distributed in a small interval around the ideal values 99.6094 and 33.4635 (shown by the horizontal lines) respectively. Hence the proposed image cipher shows extreme sensitivity on the plaintext and not vulnerable to the differential attacks like: chosen plaintext, known plaintext and adaptive chosen plaintext attacks.

## 7. The key space analysis

The secret key in the proposed image encryption technique is divided into two parts. First part consists of a set of three floating point numbers and four integers $(x_0, y_0, K, NS, NR, seed1, seed2)$, where $x_0, y_0 \in (0, 2\pi)$, $K$ can have any real value greater than 18.0, $NS$ is an integer value preferably greater than 100, $NR$ is an integer value between 1 and 16, which is treated as no. of encryption rounds, seed1 and seed2 are integers between 0 to 255 and used as the seed values for the initiation of the substitution. The second part is a 256X256 gray image whose gray pixel values form a 256X256 Latin square. The key space (i.e., the total number of different keys that can be used in the encryption/decryption) for the proposed image encryption scheme can be calculated as follows: (i) By following the discussion in Patidar *et al.* [Patidar *et al.*, 2011] the key space for the chaotic standard map parameters is $(6.28)^3 \times 10^{42}$, (ii) total possible values of NS is $10^3$, (iii) total possible values of NR is 15, (iv) total possible values of seed1 and seed2 are 256 each and the most important the minimum possible combination of quasigroup QG image of 256X256 is $10^{101724}$. Hence the total key space of the proposed image encryption algorithm is greater than $2.434 \times 10^{101769}$.

## 8. Conclusion

In a large number of proposals on chaos based image encryption schemes, it has been observed that these schemes offer a good combination of performance and security features. In most of such schemes control parameters or initial conditions of the chaotic systems have been used as part of the secret key. However, chaotic systems offer desirable properties only for the certain ranges of such parameters/initial conditions, the key space available for such image encryption techniques is greatly reduced and opens possibilities for a brute force attack. With an objective to increase the available key space and retaining all the desirable properties offered by chaos based image encryption schemes, we have proposed a simple, efficient and robust image encryption scheme based on the combined applications of quasigroups and chaotic dynamical systems. Our analysis results show that the proposed image cipher possesses a very large key space (which is practically impossible to break through brute force attack) as well as all other desirable properties which an ideal image encryption system should have. We believe that it is one of the very first steps taken and further research will open new avenues in the area of combined application of chaos and quasigroups (Latin squares) in image encryption.

| ⊙ | 1 | 2 | 3 | 4 | 5 |
|---|---|---|---|---|---|
| 1 | 2 | 1 | 5 | 3 | 4 |
| 2 | 5 | 4 | 2 | 1 | 3 |
| 3 | 3 | 5 | 1 | 4 | 2 |
| 4 | 4 | 2 | 3 | 5 | 1 |
| 5 | 1 | 3 | 4 | 2 | 5 |

| ⊘ | 1 | 2 | 3 | 4 | 5 |
|---|---|---|---|---|---|
| 1 | 2 | 1 | 4 | 5 | 3 |
| 2 | 4 | 3 | 5 | 2 | 1 |
| 3 | 3 | 5 | 1 | 4 | 2 |
| 4 | 5 | 2 | 3 | 1 | 4 |
| 5 | 1 | 4 | 2 | 3 | 5 |

| ⊘ | 1 | 2 | 3 | 4 | 5 |
|---|---|---|---|---|---|
| 1 | 5 | 1 | 3 | 2 | 4 |
| 2 | 1 | 4 | 2 | 5 | 3 |
| 3 | 3 | 5 | 4 | 1 | 2 |
| 4 | 4 | 2 | 5 | 3 | 1 |
| 5 | 2 | 3 | 1 | 4 | 5 |

**Figure 1:** Quasigroup of order 5 and its left and right inverses

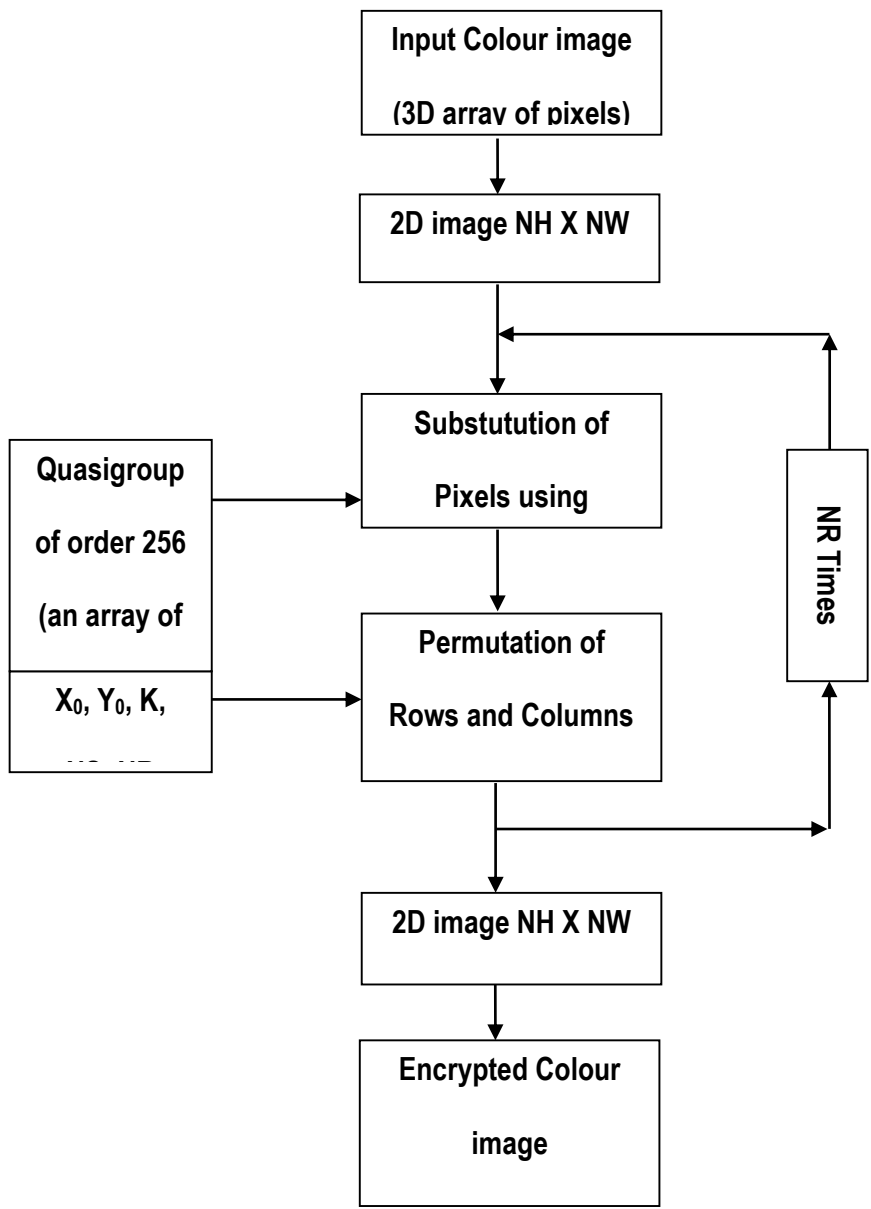

Figure 2: Block diagram of the proposed image cipher

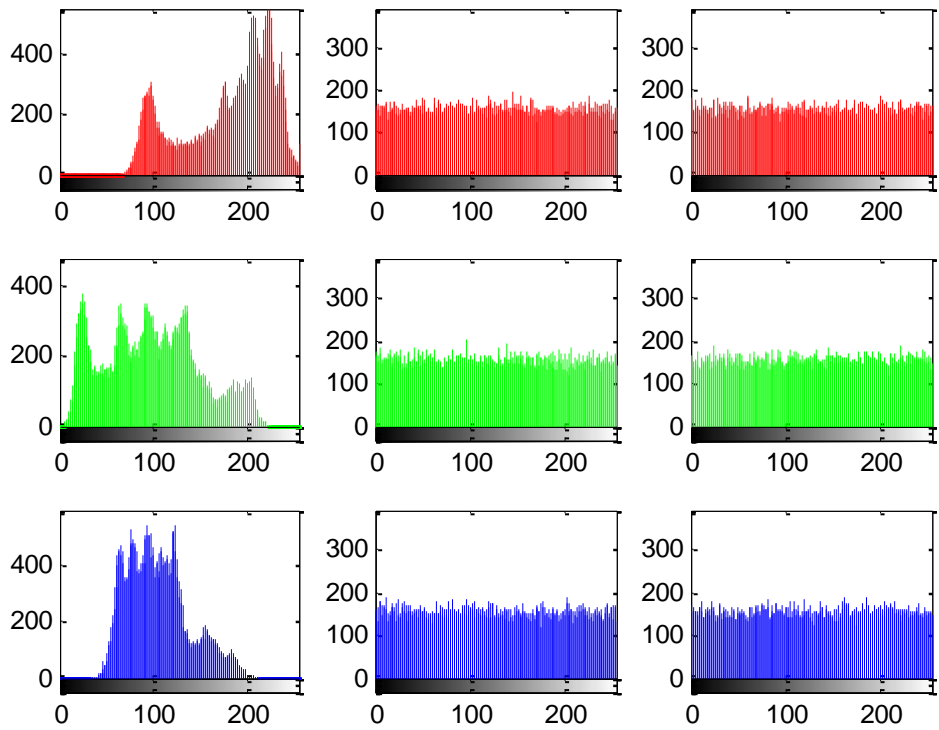

**Figure 3:** The histograms of the plain image 'Lena' (first column) and its cipher images for the first two rounds of encryption. The secret key used is given in Table 1.

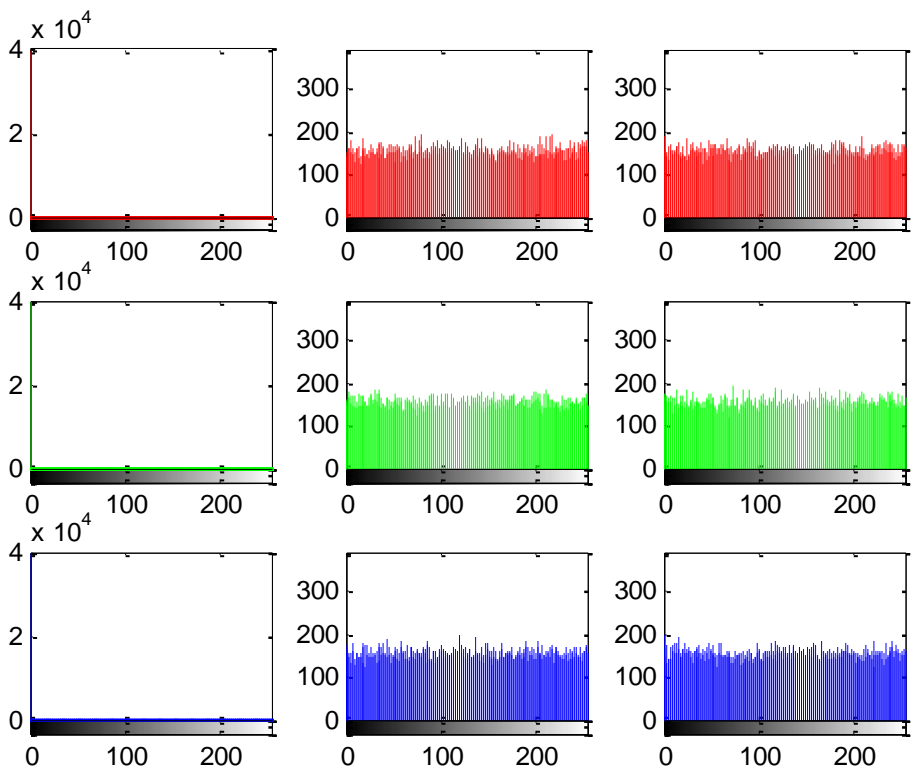

**Figure 4:** The histograms of an 'all-zero' plain image (first column) and its cipher images for the first two rounds of encryption. The secret key used is given in Table 1.

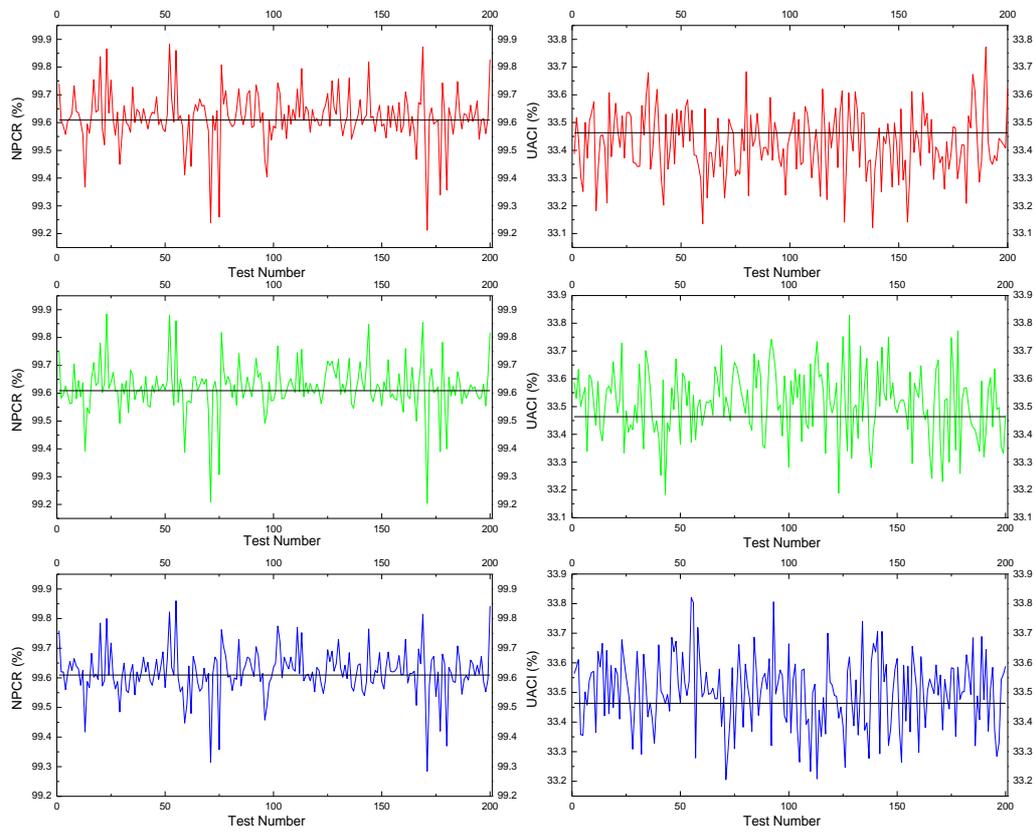

**Figure 5:** NPCR and UACI values for the plain image 'Lena'

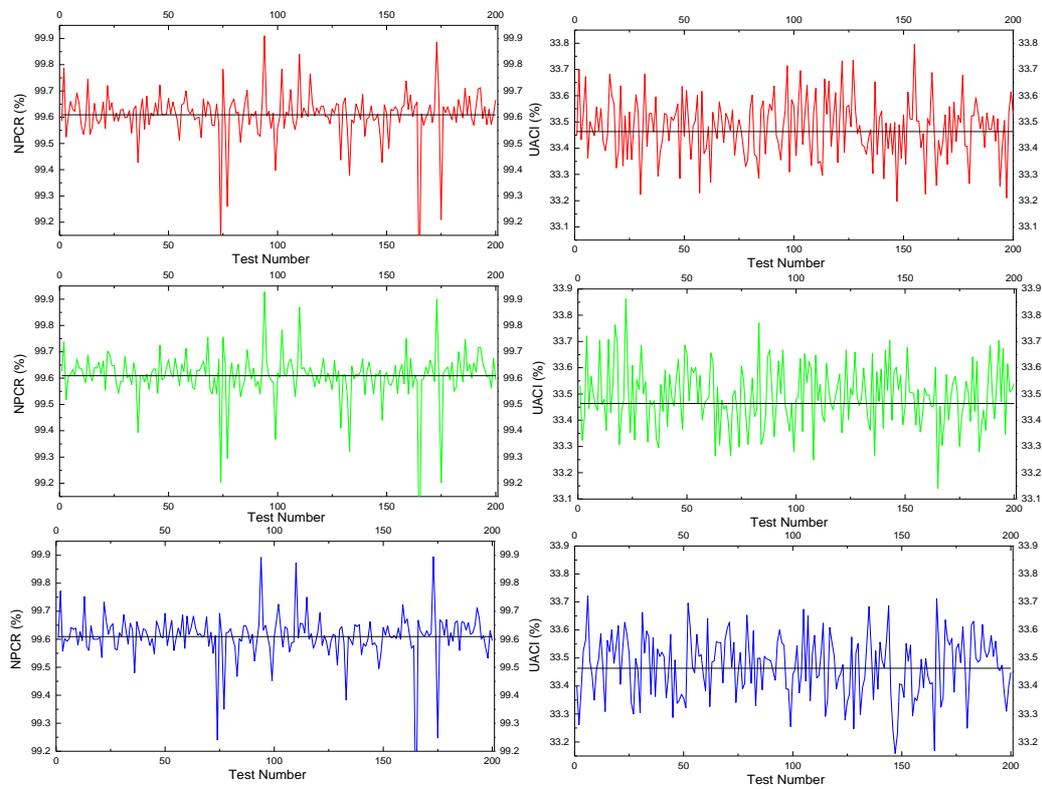

**Figure 6:** NPCR and UACI values for the "all-zero' plain image

*Revised version of this paper is accepted for publication in **Journal of Applied Nonlinear Dynamics (2017)***

Table 1: The encryption of images 'Lena' and 'All-zero'

| The secret Key | $x_0 = 2.86295319532475$<br>$y_0 = 4.56538639123458$<br>$K = 108.43745557666125$<br>$NS = 108$<br>$NR = 2$<br>$seed1 = 139$<br>$seed2 = 47$ | 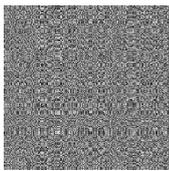<br>**Quasigroup Image 256X256** |
|---|---|---|
| **Plain Image** | 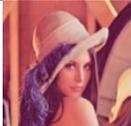 | 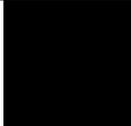 |
| **NR=1** — After Substitution | 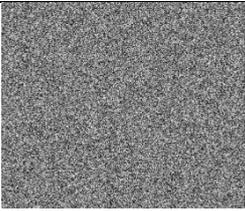 | 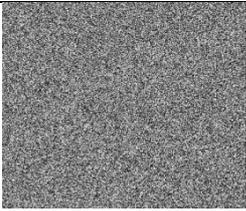 |
| **NR=1** — After Permutation | 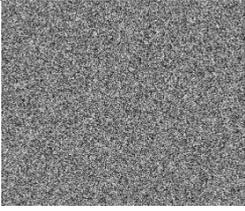 | 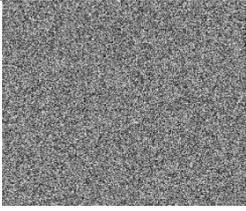 |
| **NR=1** — Encrypted Image | 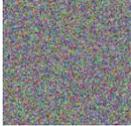 | 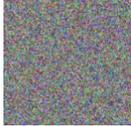 |
| **NR=2** — After Substitution | 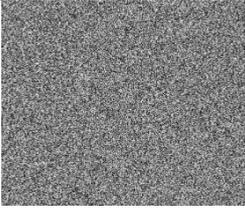 | 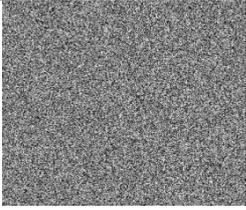 |
| **NR=2** — After Permutation | 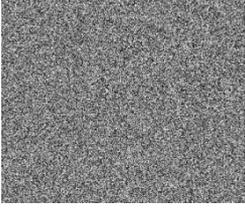 | 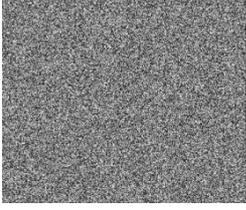 |
| **NR=2** — Encrypted Image | 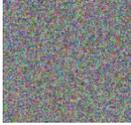 | 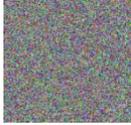 |

Table 2: Information Entropy (bits)

|  | Red | Green | Blue |
|---|---|---|---|
| Plain Image *Lena* | 7.2461 | 7.5410 | 6.9737 |
| Cipher image *Lena (NR=1)* | 7.9958 | 7.9948 | 7.9955 |
| Cipher image *Lena (NR=2)* | 7.9956 | 7.9958 | 7.9957 |
| Plain image *All Zero* | 0.0 | 0.0 | 0.0 |
| Cipher image *All zero (NR=1)* | 7.9947 | 7.9956 | 7.9953 |
| Cipher image *All zero (NR=2)* | 7.9956 | 7.9953 | 7.9954 |

Table 3: Correlation between plain and cipher images

|  |  | **Red** | **Green** | **Blue** |
|---|---|---|---|---|
| Lena | NR = 1 | | | |
|  | Red | 0.0047 | 0.0027 | 0.0044 |
|  | Green | -0.0077 | 0.0028 | -0.0033 |
|  | Blue | 0.0089 | 0.0036 | -0.0023 |
|  | NR=2 | | | |
|  | Red | -0.0024 | 0.0059 | 0.0018 |
|  | Green | -0.0052 | 0.0039 | 0.0010 |
|  | Blue | -0.0074 | 0.0028 | 0.0000 |
| All Zero | NR = 1 | | | |
|  | Red | 0.0048 | 0.0069 | 0.0036 |
|  | Green | 0.0048 | 0.0069 | 0.0036 |
|  | Blue | 0.0048 | 0.0069 | 0.0036 |
|  | NR=2 | | | |
|  | Red | 0.0051 | 0.0031 | 0.0037 |
|  | Green | 0.0051 | 0.0031 | 0.0037 |
|  | Blue | 0.0051 | 0.0031 | 0.0037 |

Table 4: Correlation between adjacent pixels

|  |  | Red | Green | Blue |
|---|---|---|---|---|
| Plain Image *Lena* | Horizontal | 0.9395 | 0.9385 | 0.8975 |
|  | Vertical | 0.9709 | 0.9706 | 0.9459 |
| Cipher image *Lena (NR=1)* | Horizontal | 0.0022 | 0.0000 | -0.0029 |
|  | Vertical | -0.0055 | 0.0033 | 0.0027 |
| Cipher image *Lena (NR=2)* | Horizontal | 0.0033 | -0.0021 | -0.0051 |
|  | Vertical | -0.0057 | 0.0001 | -0.0022 |
| Plain image *All Zero* | Horizontal | 1.0000 | 1.0000 | 1.0000 |
|  | Vertical | 1.0000 | 1.0000 | 1.0000 |
| Cipher image *All zero (NR=1)* | Horizontal | -0.0019 | 0.0026 | -0.0058 |
|  | Vertical | 0.0024 | 0.0022 | 0.0036 |
| Cipher image *All zero (NR=2)* | Horizontal | -0.0011 | 0.0023 | -0.0014 |
|  | Vertical | 0.0015 | 0.0055 | -0.0030 |

Table 5: Details of encrypted images produced for the key sensitivity analysis along with corresponding secret keys

| | Secret Key | | Cipher Image |
|---|---|---|---|
| 1. | $x_0 = 2.86295319532475$<br>$y_0 = 4.56538639123458$<br>$K = 108.43745557666125$<br>$NS = 108$<br>$NR = 2$<br>$seed1 = 139$<br>$seed2 = 47$ | 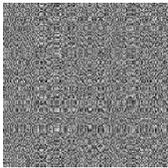<br>Quasigroup Image 256X256 | Encrypted Image |
| 2. | 1st and 2nd rows of the Quasigroup image are interchanged and the rest of the secret key remains same as in S. No. 1 | | Encrypted Image 1 |
| 3. | 255th and 256th rows of the Quasigroup image are interchanged and the rest of the secret key remains same as in S. No. 1 | | Encrypted Image 2 |
| 4. | 1st and 2nd columns of the Quasigroup image are interchanged and the rest of the secret key remains same as in S. No. 1 | | Encrypted Image 3 |
| 5. | 255th and 256th columns of the Quasigroup image are interchanged and the rest of the secret key remains same as in S. No. 1 | | Encrypted Image 4 |
| 6. | 128th and 129th rows of the Quasigroup image are interchanged and the rest of the secret key remains same as in S. No. 1 | | Encrypted Image 5 |
| 7. | 128th and 129th columns of the Quasigroup image are interchanged and the rest of the secret key remains same as in S. No. 1 | | Encrypted Image 6 |
| 8. | $Seed1$ value is changed by one unit and the rest of the secret key remains same as in S. No. 1 | | Encrypted Image 7 |
| 9. | $Seed2$ value is changed by one unit and the rest of the secret key remains same as in S. No. 1 | | Encrypted Image 8 |
| 10. | $NR$ value is changed by one unit and the rest of the secret key remains same as in S. No. 1 | | Encrypted Image 9 |
| 11. | $NS$ value is changed by one unit and the rest of the secret key remains same as in S. No. 1 | | Encrypted Image 10 |
| 12. | $K$ value is changed by $10^{-14}$ and the rest of the secret key remains same as in S. No. 1 | | Encrypted Image 11 |
| 13. | $y_0$ value is changed by $10^{-14}$ and the rest of the secret key remains same as in S. No. 1 | | Encrypted Image 12 |
| 14. | $x_0$ value is changed by $10^{-14}$ and the rest of the secret key remains same as in S. No. 1 | | Encrypted Image 13 |

Table 6: Correlation and mutual information between 'Encrypted Image' and 'Encrypted Image X' (correlate with Table 5)

| Encrypted Image X | Correlation | | | Mutual Information |
|---|---|---|---|---|
| | R | G | B | (bits) |
| Encrypted Image 1 | -0.0050 | 0.0068 | 0.0021 | 0.4505 |
| Encrypted Image 2 | -0.0058 | -0.0047 | -0.0086 | 0.4491 |
| Encrypted Image 3 | -0.0015 | -0.0037 | 0.0025 | 0.4480 |
| Encrypted Image 4 | 0.0007 | 0.0061 | 0.0048 | 0.4484 |
| Encrypted Image 5 | 0.0015 | 0.0037 | 0.0039 | 0.4470 |
| Encrypted Image 6 | 0.0033 | 0.0009 | -0.0031 | 0.4495 |
| Encrypted Image 7 | -0.0024 | -0.0009 | -0.0005 | 0.4499 |
| Encrypted Image 8 | -0.0056 | 0.0019 | -0.0075 | 0.4482 |
| Encrypted Image 9 | 0.0013 | 0.0016 | 0.0061 | 0.4517 |
| Encrypted Image 10 | -0.0023 | -0.0016 | 0.0056 | 0.4515 |
| Encrypted Image 11 | 0.0046 | -0.0066 | -0.0052 | 0.4471 |
| Encrypted Image 12 | -0.0016 | -0.0024 | -0.0010 | 0.4478 |
| Encrypted Image 13 | -0.0060 | 0.0089 | -0.0082 | 0.4496 |